\documentclass[%
 reprint,
 amsmath,amssymb,
 aps,prx,
 twocolumn,
 showpacs,
 superscriptaddress,
 floatfix,
 showkeys,
 reprint,
 longbibliography,]
 {revtex4-1}

\usepackage{graphicx}
\usepackage{dcolumn}
\usepackage{bm}
\usepackage{amsmath,amssymb}
\usepackage{hyperref}
\usepackage{natbib}
\usepackage{xcolor}
\usepackage[normalem]{ulem}
\usepackage{booktabs}
\usepackage{tabularx}
\usepackage[english]{babel}
\usepackage[utf8]{inputenc}
\usepackage{enumerate}
\usepackage{newunicodechar}
\usepackage{lipsum}

\usepackage[font=small,format=plain,labelfont=bf,
singlelinecheck=false,justification=raggedright,
labelsep=space,figurename=Fig.]{caption} 

\begin{document}

\title{Electric-field driven stability control of skyrmions in an ultrathin transition-metal film}

\author{Souvik Paul}
\email{s.paul@fz-juelich.de;paul@physik.uni-kiel.de}
\affiliation{Peter Gr$\ddot{u}$nberg Institut and Institute for Advanced Simulation, Forschungszentrum J$\ddot{u}$lich and JARA, 52425 J$\ddot{u}$lich, Germany.}
\affiliation{Institute of Theoretical Physics and Astrophysics, Christian-Albrechts-Universit$\ddot{a}$t zu Kiel, Leibnizstrasse 15, 24098 Kiel, Germany}

\author{Stefan Heinze}
\email{heinze@physik.uni-kiel.de}
\affiliation{Institute of Theoretical Physics and Astrophysics, Christian-Albrechts-Universit$\ddot{a}$t zu Kiel, Leibnizstrasse 15, 24098 Kiel, Germany}
\date{\today}

\begin{abstract}
To realize 
future spintronic applications with magnetic skyrmions -- topologically nontrivial swirling spin structures -- it is essential to achieve efficient writing and deleting capabilities of these quasi-particles. Electric-field assisted nucleation and annihilation is a promising route, however, the understanding of the underlying microscopic mechanisms is still limited. Here, we show how
the stability of individual magnetic skyrmions in an ultrathin transition-metal film can be controlled via external electric fields. We demonstrate based on 
density functional theory that it is important to consider the changes of all interactions with electric field, i.e., the pair-wise exchange, the Dzyaloshinskii-Moriya interaction, the magnetocrystalline anisotropy energy, and the higher-order exchange interactions. The energy barriers for electric-field assisted skyrmion writing and deleting obtained via atomistic spin simulations vary by up to a factor of three more than the variations of the interactions calculated from first-principles.  
This surprising effect originates from the electric-field dependent size of metastable skyrmions at a fixed magnetic field. 
The large changes of lifetimes allow the possibility of electric-field assisted thermally activated
writing and deleting of skyrmions.
\end{abstract}

\maketitle
\noindent{\large{\textbf{Introduction}}}\par
Magnetic skyrmions \cite{Bogdanov1994,tokura13} show great promises as information carriers in future magnetic memory, logic devices, and neuromorphic computing due to their nanoscale size and ultralow current-driven manipulation, achieved by spin transfer torque (STT) \cite{tokura13,tomasello14,zhou14,junichi13,sampaio13,fert13} and spin orbit torque \cite{alex2009,miron2010,miron2011,liu2012,woo2017,montoya2018,callum2020}. However, the main drawback of these techniques to realize a low-energy-dissipation device is Joule heating, which destabilizes the skyrmionic bits. The electric-field-induced manipulation offers an efficient route for creating, deleting and controlling skyrmions avoiding the heating problem. The energy dissipation can be reduced by a factor of 100 by using the electric field as compared to STT \cite{matsukura15}. Another important benefit of using an electric field is that it can be applied locally and it does not displace the skyrmionic bits, which is desirable for encoding information at a particular position.

In spite of recognizing the potential of electric-field-induced manipulation, only a few experimental studies have been reported on the electric-field-induced switching of skyrmions and skyrmion bubbles in transition-metal multilayers \cite{hsu17,schott17,ma18,titiksha18} and recently in multiferroic heterostructures \cite{Wang2020,Ba2021}. Theoretical studies have addressed either the variation of magnetic anisotropy \cite{upadhyaya15,fook16,nakatani16} or the Dzyaloshinskii-Moriya interaction (DMI) \cite{yang2018,titiksha18,desplat2021} directly by the electric field or indirectly due to the electric-field-induced strain \cite{Wang2020,Ba2021}. However, it is the interplay of the exchange interaction, the DMI, and the anisotropy which is responsible for the properties of magnetic skyrmions \cite{Bogdanov1994}. Therefore, in a study on the influence of the electric field, one needs to take the variation of all these interactions into account.  Moreover, the importance of higher-order exchange interactions (HOI) beyond the conventional Heisenberg pair-wise exchange mechanism in ultrathin films for the stability of skyrmions has recently been demonstrated \cite{hoi2020}.

Here, we investigate the stability of isolated magnetic skyrmions in an ultrathin film from first-principles electronic structures theory. We take an atomic Fe/Rh bilayer on the Re(0001) surface, Fe/Rh/Re(0001), as a model system, which is capable of stabilizing isolated skyrmions at external magnetic fields in absence of an electric field \cite{hoi2020}. From density functional theory (DFT) calculations, we find that the Heisenberg pair-wise exchange interactions vary by about 15$\%$, the DMI by only 8$\%$, while the magnetocrystalline anisotropy energy (MAE) varies by about 30$\%$ for an electric field difference of 1~V/{\AA}. Among the HOI, only the four-site four spin interaction shows a significant variation of about 20$\%$ for an electric field difference of 1~V/{\AA}. We analyze the changes of the magnetic interactions based
on the spin-dependent screening of the electric field at the surface.

We study the formation and collapse of individual skyrmions under electric fields by atomistic spin simulations using the DFT parameters. The energy barriers preventing the collapse of individual skyrmions varies by about 60$\%$ and the barriers for skyrmion creation by about 30$\%$ for an electric difference of 1~V/{\AA}. 
The enhanced energy barriers with respect to the variation of the magnetic interactions can be explained by a shift of the critical magnetic field by about 0.6~T of the phase boundary between the skyrmion and field-polarized state. Thereby, the skyrmion radius varies significantly with electric field at a given magnetic field value which leads to the large change of the energy barriers and thereby of the lifetimes which makes electric-field driven switching feasible.\\

\begin{figure*}[!htbp]
	\includegraphics[scale=0.85]{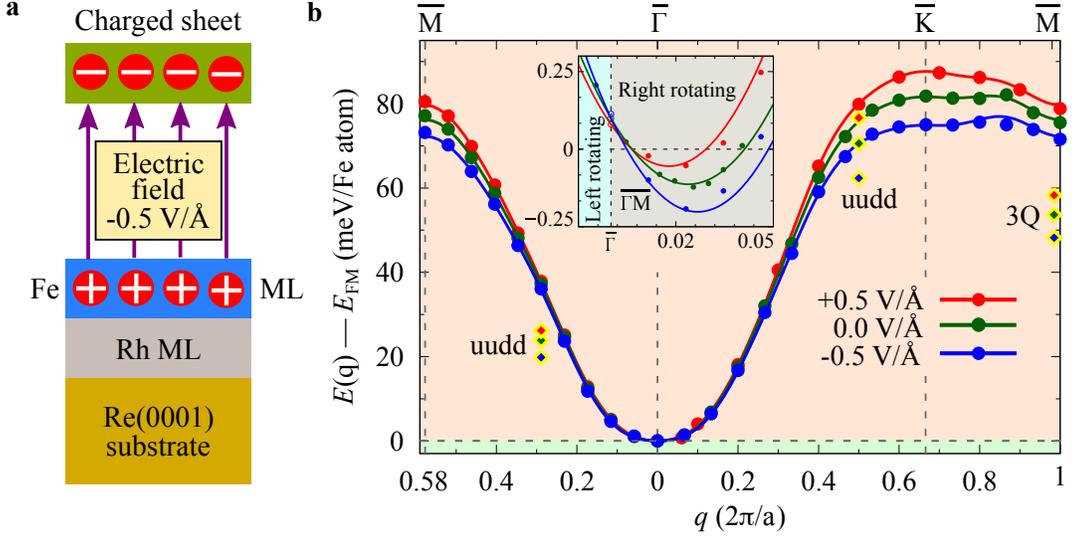}
	\centering
	\caption{\textbf{Effect of electric fields on the spin spiral energy dispersion and multi-$Q$ states. a} Illustration of the Fe monolayer (ML) on a Rh ML on the Re(0001) surface, denoted as Fe/Rh/Re(0001), exposed to a perpendicular uniform electric field. In the DFT calculation, the electric field is created by a charged sheet located at 5.3~{\AA} above the Fe layer. The direction of electric field lines for $\mathcal{E}$= $-$0.5 V/{\AA} is shown. {\textbf {b}} Energy dispersion $E(\mathbf{q})$ of flat spin spirals along two high symmetry directions ($\overline{\Gamma \mathrm{KM}}$ and $\overline{\Gamma \mathrm{M}}$) without SOC at $\mathcal{E}$= $+$0.5 V/\AA\ (red), $\mathcal{E}$= 0.0 V/\AA\ (green), and $\mathcal{E}$= $-$0.5 V/\AA\ (blue). The filled circles represent DFT data and the solid lines are fits to the Heisenberg model. The filled diamonds, highlighted by the yellow border, represent the multi-$Q$ (3$Q$ and $uudd$) states without SOC, which are shown at the $\textbf{q}$ points of the corresponding single-$Q$ states (see Fig.~\ref{fig:f6} for spin structures). Inset shows $E(\mathbf{q})$ of flat cycloidal spin spirals including DMI and MAE along $\overline{\Gamma \mathrm{M}}$ for zero and two finite electric fields. The filled circles represent DFT data and the solid lines are fits to the Heisenberg plus DMI spin model 
(see text for details). 
Grey (cyan) shaded region indicates right (left) rotating spin spirals. Note that the MAE shifts $E(\mathbf{q})$ by $K$/2 with respect to the FM state.} 
	\label{fig:f1}
\end{figure*}

\noindent{\large{\textbf{Results}}}\par

\noindent{\textbf{First-principles calculations.}}
Fig.~\ref{fig:f1}a shows the setup used to perform the DFT calculations including an electric field based on the {\tt FLEUR} code~\cite{fleur} (see methods for computational details). The electric field is introduced by placing a charged sheet in the vacuum region of the Fe/Rh/Re(0001) film \cite{weinert2009,oba15}. Charge neutrality of the whole system is maintained by adding or removing the same amount of opposite charge to or from the film. In this way, a uniform electric field perpendicular to the film plane is generated. For the electric field strength, we chose values of $\mathcal{E}$= $\pm$0.5 V/\AA\ as in the experimental work of Ref.~\cite{hsu17}, which demonstrated switching of skyrmions in Fe films at these electric field strengths.

We first discuss the energy dispersion $E(\mathbf{q})$ of homogeneous flat spin spirals without spin-orbit coupling (SOC) obtained via DFT for Fe/Rh/Re(0001) along two high-symmetric directions of the two-dimensional Brillouin zone (2DBZ) (Fig.~\ref{fig:f1}b). The magnetic moment $\mathbf{M}_i$ at lattice site $\mathbf{R}_i$, of a spin spiral is given by $\mathbf{M}_i=M (\cos{\mathbf{q}\mathbf{R}_i},\sin{\mathbf{q}\mathbf{R}_i},0)$, where $M$ is the size of the moment (about $2.9 \mu_{\rm B}$ per Fe atom) and $\mathbf{q}$ is the spin spiral vector. At $\mathcal{E}$= $0$, the ferromagnetic (FM) state ($\overline{\Gamma}$ point) is energetically lowest. The N\'eel state with an angle of 120$^\circ$ between adjacent spins ($\overline{\mathrm K}$ point) and the row-wise antiferromagnetic (AFM) state ($\overline{\mathrm M}$ point) are significantly higher in energy. The dispersion calculated for $\mathcal{E}$= $\pm$0.5 V/\AA\ shows the same trend as of the zero field. The electric field induced modification of $E(\mathbf{q})$ is not visible on this scale at small $\mathbf{q}$ (see Supplementary Fig.~1 for a close-up), however, it is significant at large $\mathbf{q}$ with an energy rise (drop) at 
$\mathcal{E}$$>$ $0$ ($\mathcal{E}$$<$ $0$). This kind of modification has been observed earlier in a freestanding Fe monolayer (ML) and for a Co ML on Pt(111), which can be explained by the spin spiral ($\mathbf{q}$) dependent screening of the electric field \cite{oba15}.  

\begin{table*}[!htbp]
	\centering
	\begin{ruledtabular}
		\begin{tabular}{ccccccccccccc}
			Electric field & $J_{1}$ & $J_{2}$ & $J_{3}$ & $J_{4}$ & $J_{5}$ & $J_{6}$ & $J_{7}$ & $B_{1}$ & $Y_{1}$ & $K_{1}$ & 
			$D_{\rm eff}$
			& $K$\\
			\midrule
			$+$0.5 V/\AA{} & 9.37 & $-$1.03 & 0.07 & $-$0.24 & 0.27 & $-$0.01 & $-$0.12 & $-$0.34 & 1.05 & $-$1.23 & 0.87 & $-0.16$ \\
			0.0 V/\AA{} & 8.85 & $-$0.77 & $-$0.05 & $-$0.22 & 0.27 & 0.05 & $-$0.16& $-$0.39 & 1.00 & $-$1.36 & 0.89 & $-0.20$ \\
			$-$0.5 V/\AA{} & 8.08 & $-$0.55 & 0.03 & $-$0.20 & 0.23 & 0.01 & $-$0.12& $-$0.33 & 1.00 & $-$1.53 & 0.94 & $-0.22$\\
		\end{tabular}
	\end{ruledtabular}
	\caption{{\textbf{Variation of interaction constants with electric fields.}} Constants for the $i$-th nearest-neighbor exchange ($J_{i}$), biquadratic exchange ($B_{1}$), three-site four spin exchange ($Y_{1}$), four-site four spin exchange ($K_{1}$), effective DMI ($D_{\rm eff}$), and the MAE ($K$) obtained via DFT for Fe/Rh/Re(0001) at zero and two finite electric fields. The positive sign of $D_{\rm eff}$ indicates a clockwise rotational sense and negative value of $K$ indicates an in-plane easy magnetization axis (easy plane). Data of $\mathcal{E}$= 0.0 V/{\AA} are taken from Ref.~\cite{hoi2020}. Note that the first three exchange interactions are modified according to Eqs.~(8a-c). For more details, see Supplementary Table 1. All energies are given in meV.}
	\label{tab:table1}
\end{table*}

SOC adds two contributions: the MAE and the DMI (inset of Fig.~\ref{fig:f1}b). The easy magnetization axis of the Fe/Rh bilayer is in the film plane (easy plane) and the value of the MAE is $K$= $-0.2$~meV/Fe atom that leads to an energy offset of $K/2$ for spin spirals with respect to the FM state ($\overline{\Gamma}$ point). The MAE changes only slightly upon applying an electric field which is barely visible in the inset. The DMI arises due to breaking of the inversion symmetry at the surface and here it favors cycloidal spin spirals with a clockwise rotational sense (inset of Fig.~\ref{fig:f1}b). For $\mathcal{E}$= $+$0.5 V/\AA, the spin spiral energy minimum is $-$0.05 meV/Fe below the FM state and the spiral exhibits a pitch of 18.4~nm. For $\mathcal{E}$= $-$0.5 V/\AA, the minimum is $-$0.20 meV/Fe lower and the pitch significantly decreases to 10.2~nm. Note that the field-induced change of the energy minimum by 0.15~meV/Fe atom is significant since the Zeeman energy due to an applied magnetic field of 1~T amounts to about 0.2~meV/Fe atom.

Since higher-order interactions (HOI) beyond pair-wise Heisenberg type exchange can be important for the magnetic ground state in transition-metal films \cite{pkurz,rhfeir111,3spin,spethmann2020,hoffmann,li2020} as well as for the stability of individual skyrmions \cite{hoi2020} and skyrmion lattices \cite{tf1}, we have also calculated their variation upon applying electric fields. To calculate the biquadratic interaction as well as the three-site and the four-site four spin interactions \cite{hoffmann}, we consider three multi-$Q$ states: the two collinear up-up-down-down ($uudd$) states \cite{hardrat} and a three-dimensional noncollinear 3Q-state \cite{pkurz} (see methods for details). As seen in Fig.~\ref{fig:f1}b, the multi-$Q$ states, calculated at zero and two finite electric fields, are lower in energy compared to the corresponding spin spiral states (for energy differences see Supplementary Table 2). The energies of the multi-Q states shift with field in a similar way as the energy dispersion, i.e., their energy increases (decreases) for positive (negative) electric field.

The total energy from DFT calculations are used to parametrize an atomistic spin model which is given by:

\begin{gather} \label{eq:hamiltonian}
\mathcal{H}=- \sum_{ij} J_{ij} (\textbf{m}_{i} \cdot \textbf{m}_{j}) - \sum_{<ij>} \textbf{D}_{ij} \cdot (\textbf{m}_{i}\times\textbf{m}_{j}) \nonumber \\ - \sum_{i} K(m^{z}_{i})^2 - \sum_{i} \mu_{s} \textbf{B} \cdot \textbf{m}_{i} - B_{1} \sum_{<ij>} (\textbf{m}_{i} \cdot \textbf{m}_{j})^2 \nonumber \\ -2~Y_{1} \sum_{<ijk>} (\textbf{m}_{i} \cdot \textbf{m}_{j}) (\textbf{m}_{j} \cdot \textbf{m}_{k}) \nonumber \\ - K_{1} \sum_{<ijkl>} (\textbf{m}_{i} \cdot \textbf{m}_{j}) (\textbf{m}_{k} \cdot \textbf{m}_{l}) +(\textbf{m}_{i} \cdot \textbf{m}_{l}) (\textbf{m}_{j} \cdot \textbf{m}_{k}) \nonumber \\ -(\textbf{m}_{i} \cdot \textbf{m}_{k}) (\textbf{m}_{j} \cdot \textbf{m}_{l})  
\end{gather}
where the magnetic moment of Fe at site $i$ is denoted by $\textbf{M}_{i}$ and $\textbf{m}_{i}$= $\textbf{M}_{i}/M_{i}$. $J_{ij}$, $\textbf{D}_{ij}$, $\mu_{s}$ and $K$ denote the pair-wise exchange constants, the DMI vectors, the magnetic moment and the MAE constant, respectively. $B_{1}$ is the biquadratic constant, $Y_{1}$ and  $K_{1}$ are the three-site and four-site four spin constant, respectively. The higher-order interactions are taken into account in nearest-neighbor approximation, 
indicated in the summation by $<..>$,
since they arise from fourth-order perturbation theory \cite{hoffmann}. The DMI has been calculated in the effective nearest-neighbor approximation, i.e., determined from the slope of the energy contribution due to SOC close to $\mathbf{q}=0$~\cite{Malottki2017a}.

The DFT values of the interaction parameters are given in table \ref{tab:table1} for zero and two finite electric fields. At $\mathcal{E}$= $+0.5$ V/{\AA}, the ferromagnetic nearest-neighbor exchange constant, $J_1$, is enhanced by about 6$\%$, while at $\mathcal{E}$= $-0.5$ V/{\AA}, it decreases by about 9$\%$ with respect to zero electric field. The absolute change of $J_1$ amounts to about 1.3~meV for a field change by 1~V/{\AA}, which is similar to the value of 1.2~meV, reported from a DFT study for a Co ML on Pt(111) \cite{oba15}. $J_2$ becomes more (less) antiferromagnetic at $\mathcal{E}$$>$ $0$ ($\mathcal{E}$$<$ $0$) with a variation almost linear with electric field. The pair-wise exchange constants beyond second neighbors remain fairly constant with electric fields. Among the HOI, only the four-site four spin interaction, $K_1$, varies significantly with electric field. It shows a decrease of about 10$\%$ at $\mathcal{E}$= $+0.5$ V/{\AA} and an increase of about 13$\%$ at $\mathcal{E}$= $-0.5$~V/{\AA}.

The effective nearest-neighbor DMI constant, $D_{\rm eff}$, varies by only 0.07~meV upon changing the electric field by 1~V/{\AA}, which amounts to a relative change of $\Delta D_{\rm eff}/D_{\rm eff} \approx 0.08$. The field-induced variation of the DMI is still twice larger than the previously reported value for a MgO/Co/Pt trilayer \cite{yang2018}. The MAE changes by 30$\%$ from $\mathcal{E}$= $+0.5$ V/{\AA} to $\mathcal{E}$= $-0.5$~V/{\AA} and it contributes 0.03~meV to the change of the spin spiral energy minimum (Fig.~\ref{fig:f1}b).   

\begin{figure*}[!htbp]
	\includegraphics[scale=1.0]{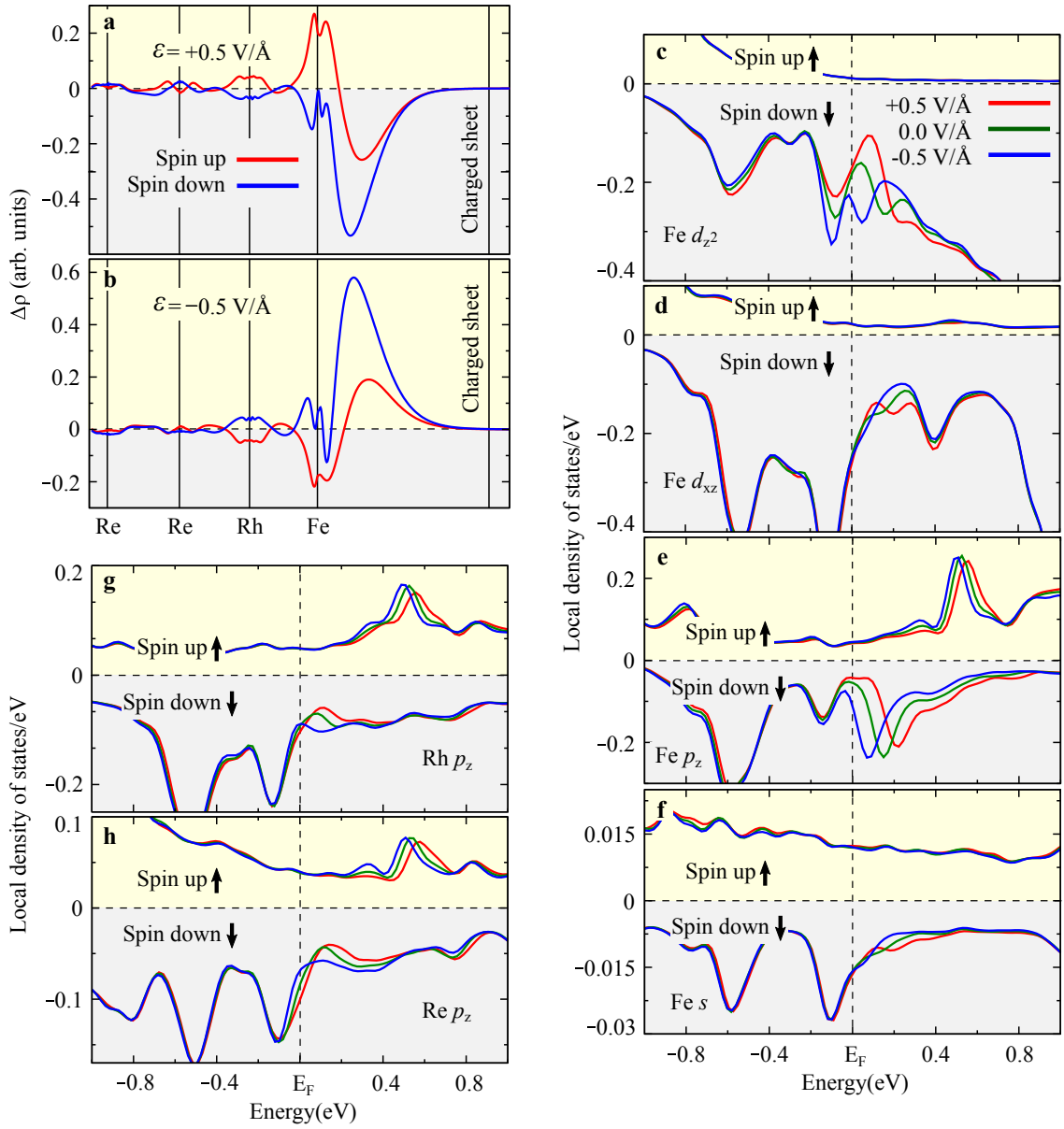}
	\centering
	\caption{{\textbf{Spin-dependent screening of electric fields at the surface. a,b}} Electric-field induced spin-dependent charge density difference $\Delta \rho_\sigma (z,\mathcal{E})$= $\rho_\sigma(z,\mathcal{E})-\rho_\sigma(z,0)$ along the $z$-direction and averaged over the film ($xy$) plane at $\mathcal{E}$= $\pm$0.5 V/{\AA}. Solid blue and red lines denote $\Delta \rho_\sigma (z,\mathcal{E})$ for spin-up ($\sigma=\uparrow$) and spin-down ($\sigma=\downarrow$) electrons, respectively. The position of the charged sheet, Fe, Rh and first two Re layers of Fe/Rh/Re(0001) film are indicated. \textbf{c-f} Spin-polarized local density of states (LDOS) of Fe $d_{z^{2}}$, $d_{xz}$, $p_{z}$, $s$, Rh $p_{z}$ and Re $p_{z}$ orbitals, respectively. Each panel displays LDOS for electric fields of $+$0.5 V/{\AA} (red), 0 V/{\AA} (green) and $-$0.5 V/{\AA} (blue). The Fermi energy is indicated as E$_{\mathrm{F}}$.} 
	\label{fig:f2}
\end{figure*}

To obtain 
insight into the electric-field induced changes of the magnetic interactions, we analyze the spin-dependent screening of the electric field in the film. The modification of the charge density, $\Delta \rho$, at positive and negative electric fields with respect to zero field is displayed for the ferromagnetic state in Fig.~\ref{fig:f2}a,b. $\Delta \rho$ is spin-dependent and therefore, it creates surface electric dipoles which screen the electric field close to the surface of the film. The direction of the electric dipole at the surface switches upon changing the sign of the electric field. We observe that $\Delta \rho$ is sizable at the Fe layer, which quickly decays to small values in the film and it displays Friedel oscillations, as expected due to screening at a metal surface \cite{mitsui2020}.

The largest charge differences, observed in front of the surface (Fig.~\ref{fig:f2}a,b), originate from the extended $p_z$ and $d_{z^{2}}$ states, which can be understood from the orbitally decomposed spin-resolved LDOS of the Fe atom (Fig.~\ref{fig:f2}c-f). The decrease (increase) of the spin down density at positive (negative) electric field in the Fe layer and the vacuum region is due to minority $p_z$ and $d_{z^{2}}$ states, just below the Fermi energy (Fig.~\ref{fig:f2}c,e). A similar field effect arises for the spin up channel in the vacuum region, which can be attributed to the majority $p_z$ states. On the other hand, the increase of the spin up charge density at the Fe atom for $\mathcal{E}>0$ 
(Fig.~\ref{fig:f2}a)
is due to an increase of the majority $d$ states as can be concluded from the change of the integrated charge density within the muffin-tin spheres (see Supplementary Table 3).

In contrast, the electric-field induced effect is very small for the Rh and Re atoms, for which only $p_{z}$ orbitals are slightly affected in the LDOS (Fig.~\ref{fig:f2}g-h). Note that the minority $p_z$ peak of Re and Rh just below the Fermi energy is strongly hybridized with the Fe $p_z$ state, as can be concluded from the same energy position and peak shape as well as from the band structures (see Supplementary Figures 2 to 5). Thereby, the $p_z$ state of Re and Rh is affected as the Fe $p_z$ state is filled or emptied due to an electric field.
 
The spin-dependent imbalance of surface charge caused by the electric field leads to a small linear change of the Fe magnetic moment with electric field, while the induced moments of Rh and Re remain almost 
unchanged (Supplementary Figure 6). Note that the Fe moment for spin spirals at zero and finite electric fields remains fairly constant upon varying $\textbf{q}$.

The increase of the nearest-neighbor exchange interaction (cf.~Table 1) in the Fe ML can be explained based on the change of the charge density. Upon increasing the electric field from $\mathcal{E}$= $-0.5$ V/{\AA} to $\mathcal{E}$= $+0.5$~V/{\AA}, there is a decrease of the total $sp$ charge by 2.5$\times 10^{-3}$ electrons and an increase of the total $d$ charge by 6.8$\times 10^{-4}$ electrons (see Supplementary Table 3). The Slater-Pauling curve relates the Curie temperature to the number of $d$ electrons in an alloy and here, we find that such an increase of $d$ electrons enhances the Curie temperature of Fe, which can be related to an enhanced nearest-neighbor exchange interaction \cite{Takahashi2007,oba15}.

The electric-field induced change of the DMI can be understood based on the electric dipoles at the interfaces of the film. The DMI constant can be related to the electric dipole moment at the interfaces of magnetic multilayers~\cite{jia2020}. Since the screening charge penetrates into the ultrathin film (Fig.~\ref{fig:f2}a,b), it contributes to the electric dipole moment. Upon changing from a positive to a negative  electric field, switching of the dipole moment causes the observed increase and decrease of the DMI strength, respectively. The Re layer is quite well screened from the electric field, therefore, its contribution to the change of the DMI is quite small despite being a $5d$ transition metal with a large SOC constant (Supplementary Figure 7). A larger effect is expected from Rh due to the non-negligible screening charge density at the Fe/Rh interface (Fig.~\ref{fig:f2}a,b). However, its contribution to the total DMI is small since Rh is a 
$4d$ transition metal. 
As a result, the total change of DMI constant is relatively small compared to that of the exchange constant. 
Nevertheless, it has a profound effect on skyrmion switching as discussed below.

The hybridization between the $p_{z}$ and $d_{z^{2}}$ states of Fe is significantly affected by the electric field. It has been shown previously that the electric-field dependent $p$-$d$ hybridization can explain the change of the magnetocrystalline anisotropy energy \cite{Nakamura2009}.  

Summarizing the DFT results, we find that the electric field influences the ground state through the variation of 
exchange interactions, DMI, and MAE, which leads to the change of the spin spiral period and the depth of the energy minimum.\\

\begin{figure}[!htbp]
	\includegraphics[scale=0.9]{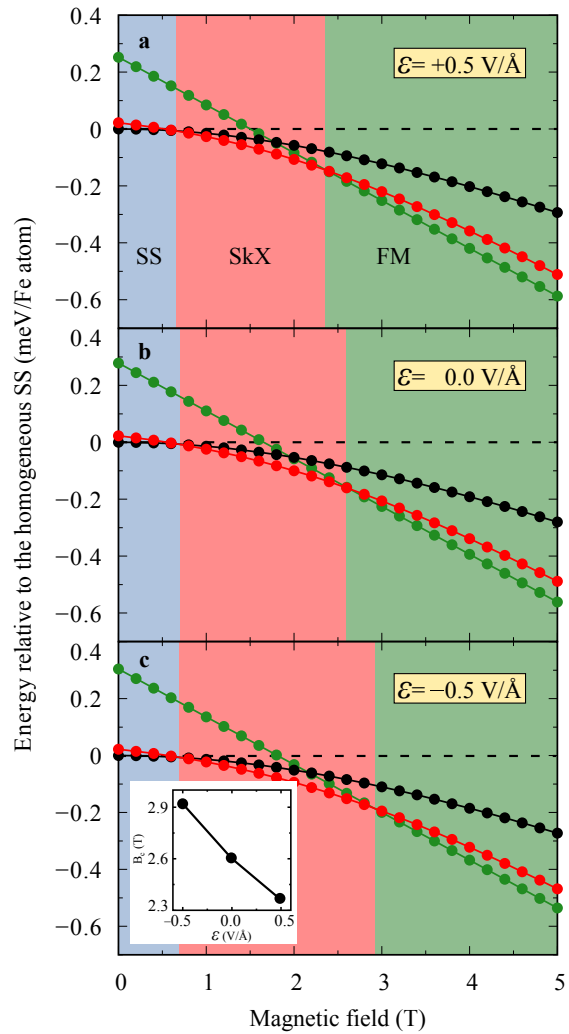}
	\centering
	\caption{\textbf{Effect of electric fields on the phase diagram.} Zero temperature phase diagram of Fe/Rh/Re(0001) at \textbf{a} $\mathcal{E}$= $+0.5$ V/\AA\ , \textbf{b} $\mathcal{E}$= $0$~V/\AA\ , and \textbf{c} $\mathcal{E}$= $-0.5$~V/\AA. Energies of the relaxed spin spiral (SS, black circles), skyrmion lattice (SkX, red circles) and field-polarized ferromagnetic (FM, green circles) states are shown with reference to the homogeneous spin spiral (dashed line). The SS, SkX and FM phases are denoted with blue, red and green background colors, respectively. Inset of \textbf{c} shows the variation of the critical field $B_{\mathrm c}$ (onset of FM phase) with the electric field. Note that the critical field is calculated where the energy of the skyrmion lattice (SkX) phase becomes equal to the FM phase.} 
	\label{fig:f3}
\end{figure}
\begin{figure}[!htbp]
	\includegraphics[scale=0.9]{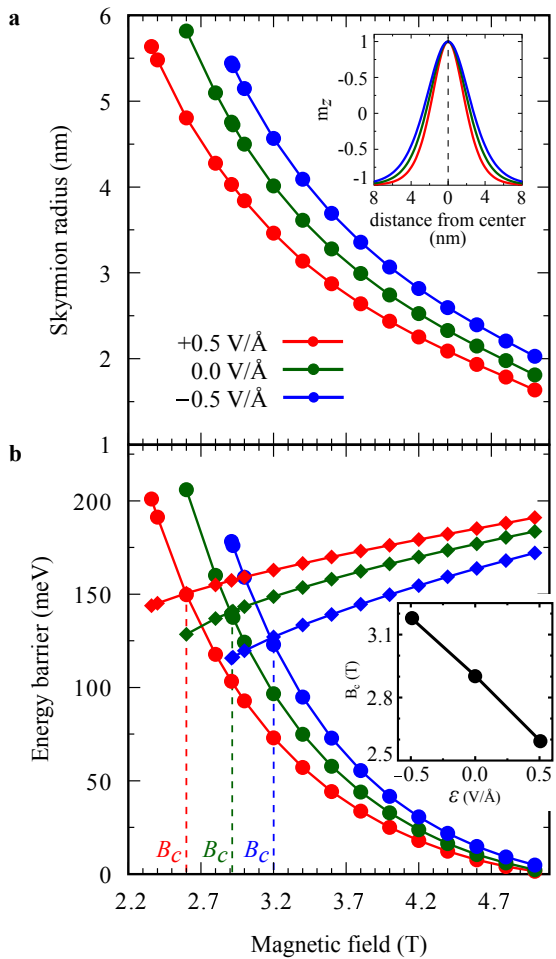}
	\centering
	\caption{\textbf{Effect of electric fields on radius and energy barriers of isolated skyrmions. a} Radius and \textbf{b} energy barriers of isolated skyrmions in Fe/Rh/Re(0001) with applied magnetic fields at $\mathcal{E}$= $+0.5$ V/\AA\ (red), $\mathcal{E}$= 0~V/\AA\ (green) and $\mathcal{E}$= $-0.5$ V/\AA\ (blue). In \textbf{b}, filled circles show collapse barriers, $\Delta E_{\rm col}$, while filled diamonds mark creation barriers, $\Delta E_{\rm crea}$. Inset of \textbf{a} shows the skyrmion profiles at $B$= $3$~T for zero and two finite electric fields and \textbf{b} shows the variation of the critical field $B_{\mathrm c}$ (onset of FM phase) with the electric field. Note that the critical field is calculated where the energy of isolated skyrmions is equal to the FM phase.} 
	\label{fig:f4}
\end{figure}

\noindent{\textbf{Atomistic spin simulations.}}
Next we show by atomistic spin dynamics simulations based on Eq.~(1) with the DFT parameters that the field-dependent spin spiral minimum allows to tune the onset of the FM phase, where magnetic skyrmions are metastable. From the zero-temperature phase diagram (Fig.~\ref{fig:f3}a-c), we find that, independent of the electric field, a spin spiral ground state occurs at zero and small magnetic fields, consistent with the spin spiral minima in Fig.~\ref{fig:f1}b. At $B$$>$ $0.7$~T, the skyrmion lattice phase becomes energetically favorable. With further increase of the magnetic field, the field-polarized 
or ferromagnetic (FM) phase becomes the lowest energy state. Note that the onset of the FM phase, i.e., the critical field $B_{\mathrm c}$, changes with electric field (inset of Fig.~\ref{fig:f3}c). The change of $B_{\mathrm c}$ (0.56~T) with the electric field (1 V/\AA) is close to the estimated value from the spin spiral energy minimum in Fig.~\ref{fig:f1}b ($\approx$ 0.75~T).   

Now we consider individual magnetic skyrmions in the field-polarized phase. We find that the skyrmion radius increases significantly for $\mathcal{E}$$<$ $0$, while it decreases for $\mathcal{E}$$>$ $0$ (Fig.~\ref{fig:f4}a). At first sight, it seems surprising that the order of skyrmion radius with the electric field is reversed with respect to the change of the spin spiral period at the energy minimum found from DFT (Fig.~\ref{fig:f1}b). However, this is due to the fact that the critical magnetic field, at which the transition from the skyrmion to the field-polarized or ferromagnetic (FM) phase occurs, also changes with the electric field. Since, in an experiment, switching of a skyrmion is performed at a fixed magnetic field \cite{tf2,hsu17}, the variation of the skyrmion radius (see inset of Fig.~\ref{fig:f4}a) strongly affects the energy barriers for skyrmion creation or annihilation as shown below \cite{footnote1}.

\begin{figure}[!t]
	\includegraphics[scale=0.837]{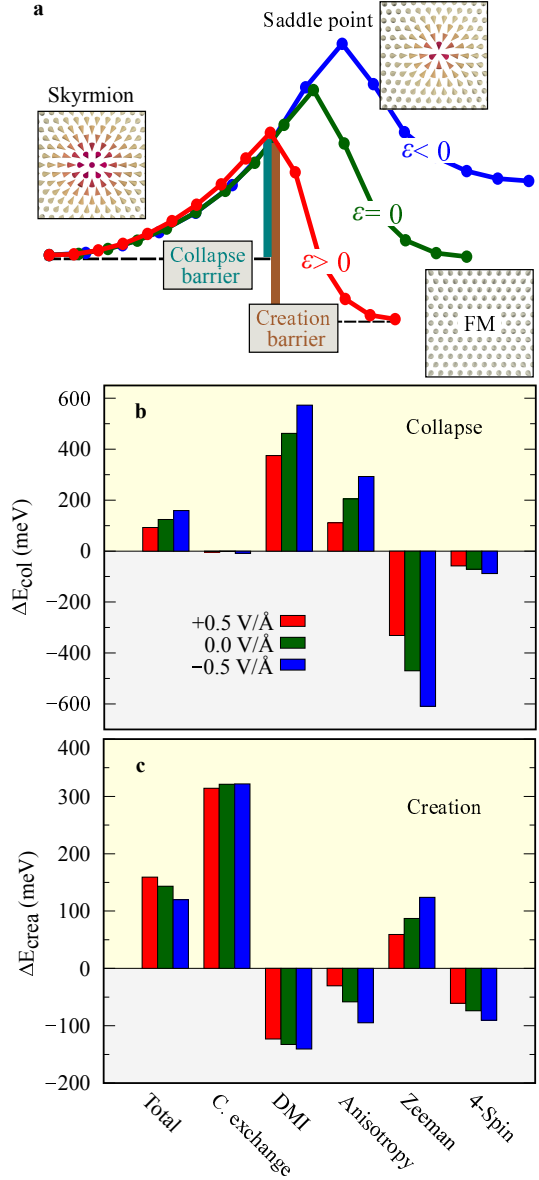}
	\centering
	\caption{\textbf{Minimum energy path, collapse and creation barriers at zero and finite electric fields.} \textbf{a} Minimum energy paths (MEP) from the initial (skyrmion) to the final (FM) state in Fe/Rh/Re(0001) at $B$= $3$~T for zero and two finite electric fields ($\mathcal{E}$= $\pm 0.5$ V/{\AA}). The energy difference between the saddle point and the skyrmion (FM) state defines the collapse (creation) barrier. Total and individual energy contributions at $B$= $3$~T for \textbf{b} the collapse barrier and \textbf{c} for the creation barrier. For energy decomposition of the full MEP, see Supplementary Fig.~8. Combined exchange (C. exchange) indicates the sum of pair-wise exchange, biquadratic and three-site four spin interactions. The four-site four spin interaction is denoted as 4-Spin for brevity. The electric fields $\mathcal{E}$= $+0.5$ V/\AA\, 0~V/\AA\ and $-0.5$ V/\AA\ are shown as red, green and blue, respectively.} 
	\label{fig:f5}
\end{figure}

We used the geodesic nudged elastic band (GNEB) method \cite{Bessarab2015} to calculate the minimum energy path (MEP) between an isolated skyrmion and the FM background 
(Fig.~\ref{fig:f5}a). We observe that the skyrmions annihilate by the well-known radial collapse mechanism ~\cite{muckel2021}. From the MEP, we find the energy barrier $\Delta E_{\rm col}$ protecting the skyrmion from collapsing into the FM state, i.e., the difference between the saddle point (SP), maximum energy point on the path, and the initial (skyrmion) state (Fig.~\ref{fig:f5}a). The creation barrier, $\Delta E_{\rm crea}$, is obtained from the difference between the SP and the final (FM) state.\\

It is apparent that $\Delta E_{\rm col}$ decreases as a function of applied magnetic field similar to the decrease of the skyrmion radius (Fig.~\ref{fig:f4}b). This is due to the fact that the energy terms which contribute to the barrier, in particular the DMI, scale with the number of spins in the skyrmion. On the other hand, $\Delta E_{\rm crea}$  displays only a small nearly linear rise with applied magnetic field (Fig.~\ref{fig:f4}b). This can be understood from the fact that it depends on the energy difference of the FM state and the saddle point. We can relate $\Delta E_{\rm crea}$ to the energy difference between the skyrmion and the FM state, $\Delta E_{\rm sk-FM}$, and the collapse barrier by $\Delta E_{\rm crea}=\Delta E_{\rm sk-FM}+\Delta E_{\rm col}$ (Fig.~\ref{fig:f5}a). As discussed above, $\Delta E_{\rm col}$ decreases with the magnetic field. However, the FM state becomes more favorable with increasing magnetic field and $\Delta E_{\rm sk-FM}$ increases (Fig.~\ref{fig:f5}a). The gradual rise of $\Delta E_{\rm crea}$ with magnetic field is due to these two opposing contributions.

The magnetic field at which both barriers are equal (green curve in Fig.~\ref{fig:f5}a) marks the transition from the isolated skyrmion to the FM phase. As seen in the inset of Fig.~\ref{fig:f4}b, this critical magnetic field shifts with the electric field. At $\mathcal{E}$= $0$ V/\AA, $B_{\rm c}$ is about 2.9~T. If one changes the electric field to $+0.5$~V/{\AA}, at a fixed magnetic field, the creation barrier is enhanced, while the collapse barrier decreases since $B_c$ shifts to a lower value (Fig.~\ref{fig:f3}a). Therefore, the FM state becomes more favorable (red curve in Fig.~\ref{fig:f5}a). For $\mathcal{E}$$<$ $0$, the opposite effect occurs since one moves into the skyrmion phase at a fixed magnetic field of 2.9~T. The skyrmion state is therefore lower (blue curve in Fig.~\ref{fig:f5}a) than the FM state and the collapse barrier rises. This explains why collapse and creation barriers show opposite trends upon variation of the electric field. Note that the critical fields in Fig.~\ref{fig:f4}b is slightly higher than the fields in Fig.~\ref{fig:f3}c. The reason is that, in the first case, the critical fields are obtained from the energy equality of the isolated skyrmion to the FM phase, and in the second case, from the energy equality of the skyrmion lattice phase to the FM phase. 

Now we quantify the energy barrier at $B$= $3$~T in terms of the magnetic interactions (Fig.~\ref{fig:f5}b,c). At this value, we find $\Delta E_{\rm col}$$\approx$ $125$ meV for zero electric field. $\Delta E_{\rm col}$ increases by $\approx 35$ meV at $\mathcal{E}$= $-0.5$ V/\AA\ (Fig.~\ref{fig:f5}b). Interestingly, the total energy barrier arises due to the increase of DMI ($\approx 110$~meV) and the MAE ($\approx 90$~meV), while both the Zeeman term ($\approx -140$~meV) and the four-site four spin interaction ($\approx -20$~meV) lead to opposite contributions and thus reduce the electric field effect \cite{footnote3}. The combined exchange interaction, i.e., the sum of the contributions from pair-wise exchange, biquadratic and three-site four spin interaction, does not contribute to the change of energy barriers. For $\mathcal{E}$= $+0.5$ V/\AA, we find a decrease of $\Delta E_{\rm col}$ by $\approx 35$ meV and
the magnetic 
interactions contribute in an analogous way.

The variation of $\Delta E_{\rm col}$ by about $\pm 30\%$ for $\mathcal{E}$= $\pm$0.5 V/{\AA} 
(Fig.~\ref{fig:f5}b)
cannot be directly understood from the electric-field induced changes of the magnetic interactions. Since the magnetic moment does not change much with the electric field, the large variation of the Zeeman energy must be directly linked to the change of the skyrmion radius with the electric field (Fig.~\ref{fig:f4}a). For $\mathcal{E}$= $\pm$0.5 V/{\AA}, the skyrmion radius varies by 1.3~nm at $B$= $3$~T. We can estimate the relative change of the number of moments in the skyrmion from $\Delta N_{\rm sk}/N_{\rm sk} = (R^2_{{\rm sk}}(\mathcal{E})-R^2_{\rm sk}(0))/R^2_{\rm sk}(0)\approx 0.3$, which closely matches the variation of the Zeeman contribution to the energy barrier $\Delta E_{\rm Zeeman}/E_{\rm Zeeman}\approx 0.3$. The electric-field dependent skyrmion radius also explains the large relative change of the DMI term, $\Delta E_{\rm DMI}/E_{\rm DMI} \approx 0.25$, and the anisotropy term, $\Delta E_{\rm MAE}/E_{\rm MAE} \approx 0.45$, which are about four times larger than the change of $D_{\rm eff}$ and $K$, respectively (Table \ref{tab:table1}). For the four-spin interaction, $K_1$, this effect is much reduced since the barrier contribution of this term depends on the saddle point structure \cite{hoi2020} which is less affected by the electric field.

The creation barrier (Fig.~\ref{fig:f5}c) varies by about 15\% for $\mathcal{E}$= $\pm$0.5 V/{\AA}, but displays an opposite trend with respect to $\Delta E_{\rm col}$, i.e., the barrier decreases for $\mathcal{E}$$<$ $0$ and rises for $\mathcal{E}$$>$$0$. As stated above, this can be explained based on the MEP (Fig.~\ref{fig:f5}a). The energy decomposition (Fig.~\ref{fig:f5}c) shows that the DMI plays a minor role, while MAE and four-spin exchange determine the electric field dependence \cite{footnote2}. The Zeeman term shows an opposite trend and again reduces the electric field effect. The scaling of anisotropy and Zeeman term with electric field is also much larger than expected from the electric-field induced changes of the magnetic interactions and is due to the change of skyrmion radius as explained above
for $\Delta E_{\rm col}$.\\

\noindent{\large{\textbf{Discussion}}}\par

Based on our results, we can discuss the electric-field assisted switching of skyrmions in a scanning tunneling microscopy (STM) experiment \cite{hsu17}. The electric current injected from the STM tip results in magnon excitations which eventually trigger the skyrmion collapse. In Ref.~\cite{muckel2021}, an effective temperature of the magnon bath due to single hot electron events was estimated and it is around $50$~K. The skyrmion collapse and creation rates calculated based on the Arrhenius law compared well with experimental values \cite{muckel2021}. Therefore, we use it to estimate the effect of the electric field on the switching.

The Arrhenius law is given by $\tau$= $\tau_0 \exp{(\Delta E_{\rm col}/k_{\rm B}T)}$. Since the lifetime $\tau$ is dominated by the energy barrier $\Delta E_{\rm col}$ at low temperatures, we neglect the variation of the prefactor
$\tau_0$. Then, the ratio of lifetimes at $B$= $3$~T is given by $\tau(\mathcal{E}$= $+0.5~{\rm V/{\AA}})/\tau(\mathcal{E}$= $0)$= $\exp{(- 35~{\rm meV}/k_{\rm B}T)}$. This leads to a change of the skyrmion lifetime by a factor of about $2$$\times$$10^{-4}$ at a temperature of $T$= $50$~K. Therefore, a significant effect in deleting individual skyrmions can be obtained by a local electric field in the tunnel junction. A similar estimate for the creation leads to a factor of $8$$\times$$10^{-3}$ which shows that writing of skyrmions 
can also be greatly manipulated by an electric field.

We have demonstrated that the stability of isolated skyrmions in an ultrathin film can be changed significantly by external electric field. The electric-field induced changes of the magnetic interactions lead to a shift of the critical magnetic field for the onset of the field-polarized phase, which exhibits metastable skyrmions. As a consequence, changes in the creation and collapse barriers are much larger than expected from the variations of the interactions which lead to the
possibility of writing and deleting skyrmions. Our study shows that it is indispensable to consider all magnetic interactions to evaluate the electric-field effect for skyrmion stability.\\                
 
\noindent{\large{\textbf{Methods}}}\par

\noindent{{\textbf{Density functional theory calculations.}}
All DFT calculations with applied electric fields were performed using the \textsc{fleur} code \cite{fleur}. We relaxed the top three layers of Fe/Rh/Re(0001), i.e., the Fe, the Rh and the first Re layer, along the $z$-direction by minimizing the atomic force on each atom down to values of less than 0.04 eV/{\AA} in the presence of $\mathcal{E}$= $\pm$0.5 V/\AA. We chose the generalized gradient approximation (GGA) exchange-correlation functional as parametrized by Perdew, Burke and Ernzerhof \cite{ggapbe}, 66 $k$ points in two-dimensional Brillouin zone (2DBZ) and $k_{max}$= $4.0$ a.u.$^{-1}$ for relaxation. We found that there were almost no changes  (below 0.01~a.u.) of the top three inter-layer distances compared to the zero field values given in Ref.~\cite{paul2020}. Therefore, we take the zero electric field interlayer distances of Ref.~\cite{paul2020} for calculations including electric fields. Similar to our result, no electric-field induced relaxation was observed in a Co ML on the Pt(111) surface \cite{oba15}.

To check the existence of a noncollinear ground state and to extract the Heisenberg pair-wise exchange parameters, we calculated the energy dispersion of homogeneous flat spin spirals of the form $\mathbf{M}_i=M (\cos{\mathbf{q}\mathbf{R}_i},\sin{\mathbf{q}\mathbf{R}_i},0)$, where $\mathbf{M}_i$ is the magnetic moment at lattice site $\mathbf{R}_i$ and $\mathbf{q}$ is the wave vector in 2DBZ \cite{kurz2004}. We used the generalized Bloch theorem \cite{gbt} to compute spin spiral energies within the chemical unit cell. To study the surface, we chose an asymmetric film of two atomic overlayers on nine Re substrate layers. Since we used an asymmetric film to compute the magnetic interactions and our interest is on the electric-field induced changes in magnetism, we only apply electric fields perpendicular to the Fe surface. To be consistent with the zero electric field spin spiral calculations~\cite{paul2020}, we used the local density approximation 
(LDA) exchange-correlation functional form given by Vosko, Wilk and Nusair \cite{vwn}, a dense mesh of 44$\times$44 $k$-points in the full 2DBZ and $k_{max}= 4.0$ a.u.$^{-1}$.   

The DMI was computed within the first-order perturbation theory \cite{heide} on the self-consistent spin spiral state. The MAE was calculated 
self-consistently as described in Ref.~\cite{Li1990} starting from a self-consistent scalar-relativistic density. 
To obtain an accurate value of the MAE, we varied the substrate layers from 13 to 17 of the asymmetric film.\\

\noindent{\textbf{Higher-order exchange interactions from DFT.}}
The fourth-order perturbative expansion in the hopping parameter over the Coulomb interaction parameter of the Hubbard model \cite{hubbard1} leads to a two-site four spin (biquadratic, $B_{1}$), a three-site four spin ($Y_{1}$) \cite{hubbard2,hubbard3} and a four-site four spin ($K_{1}$) \cite{hoffmann} term (see Eq. (1)). These terms arise due to hopping of the electron among two (two-site four spin or biquadratic), three (three-site four spin) and four (four-site four spin) lattice sites. We evaluate these three higher-order interaction constants directly from DFT within nearest-neighbor approximation.

\begin{figure*}[!htbp]
	\includegraphics[scale=0.8]{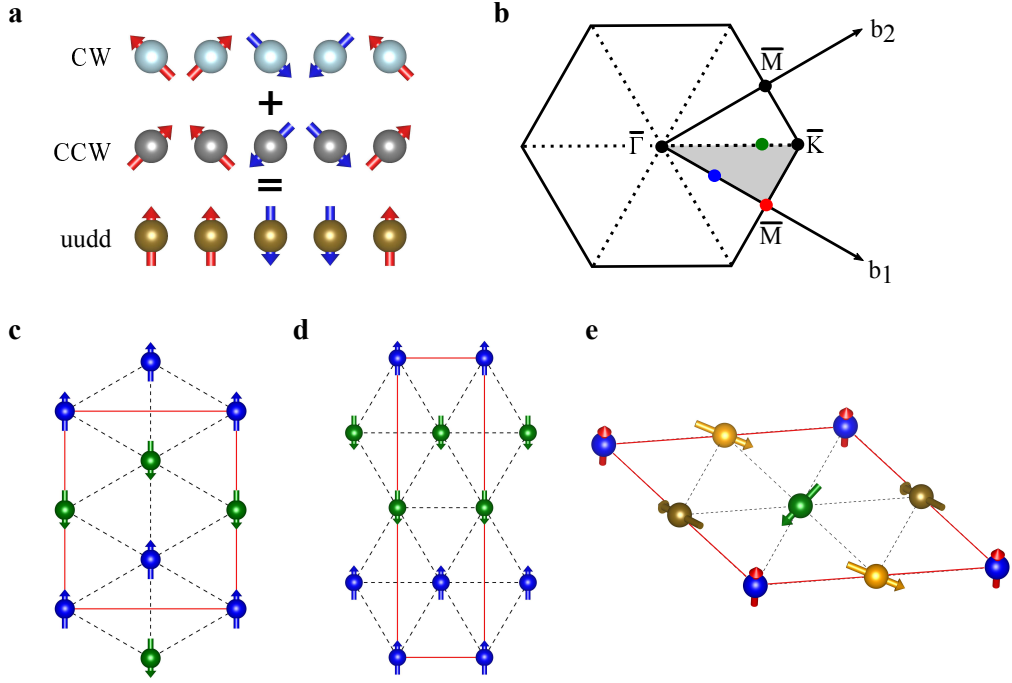}
	\centering
	\caption{\textbf{Formation and spin structures of the multi-$Q$ states. a} Superposition of two 90$^\circ$ spin spirals (1$Q$ states) with clockwise (CW) and counterclockwise (CCW) rotational sense results in an $uudd$ state ($2Q$ state). \textbf{b} 2D Brillouin zone of hexagonal lattice and reciprocal vectors $\textbf{b$_{1}$}$ and $\textbf{b$_{2}$}$. Two high-symmetry directions $\overline{\Gamma \mathrm{K}}$ and $\overline{\Gamma \mathrm{M}}$ are shown. The $\textbf{q}$ vectors corresponding to the $uudd$ state along $\overline{\Gamma \mathrm{K}}$ (green circle) and along $\overline{\Gamma \mathrm{M}}$ (blue circle) as well as the 3$Q$ state at the $\overline{\mathrm{M}}$ point (red circle) are shown. \textbf{c} Spin structure of the $uudd$ state along $\overline{\Gamma \mathrm{K}}$, which is formed by a superposition of two 90$^\circ$ spin spirals at $\textbf{q}$= $\pm (3/4) \overline{\Gamma \mathrm{K}}$, \textbf{d} spin structure of the $uudd$ state along $\overline{\Gamma \mathrm{M}}$, which is formed by a superposition of two 90$^\circ$ spin spirals at $\textbf{q}$= $\pm (1/2) \overline{\Gamma \mathrm{M}}$ and \textbf{e} spin structure of the 3$Q$ state, which is formed by a superpostion of three spin spirals at the $\overline{\mathrm{M}}$ point. Note that the spin structure of the $uudd$ states is collinear, whereas it is non-collinear for the 3$Q$ state.}
	\label{fig:f6}
\end{figure*}

To compute the constants, we consider three multi-$Q$ states (Fig. \ref{fig:f6}c-e). The multi-$Q$ state is constructed from the superposition of spin spirals (single-$Q$) related to symmetry equivalent $\textbf{q}$ vectors of the 2DBZ 
(Fig.~\ref{fig:f6}a,b). The single-$Q$ and multi-$Q$ states are energetically degenerate within the Heisenberg pair-wise spin model. The degeneracy is lifted when the mentioned three higher-order interactions are taken into account which provides a way to compute their strength.

We consider two collinear $uudd$ or 2$Q$ states \cite{hardrat} and a three-dimensional noncollinear 3$Q$ state \cite{pkurz} to determine the three higher-order exchange constants (for spin structures see Fig.~\ref{fig:f6}c-e). The three higher-order exchange interaction constants without SOC are determined from the energy differences of the single-$Q$ and multi-$Q$ states using the following equations:
\begin{gather} 
B_{1}=\frac{3}{32} \Delta E_{\overline{\mathrm{M}}}^{3Q} - \frac{1}{8} \Delta E_{\overline{\mathrm{M}}/2}^{uudd} \label{eq6} \\
Y_{1}=\frac{1}{8}(\Delta E_{3\overline{\mathrm{K}}/4}^{uudd} - \Delta E_{\overline{\mathrm{M}}/2}^{uudd}) \label{eq7} \\
K_{1}=\frac{3}{64} \Delta E_{\overline{\mathrm{M}}}^{3Q} + \frac{1}{16} \Delta E_{3{\overline{\mathrm{K}}}/4}^{uudd} \label{eq8}
\end{gather}
where the total energy differences between multi-$Q$ and corresponding single-$Q$ (spin spiral) states given by
\begin{gather}
\Delta E_{\overline{\mathrm{M}}}^{3Q}= E_{\overline{\mathrm{M}}}^{3Q} - E_{\overline{\mathrm{M}}}^{SS}	 \label{eq4a} \\
\Delta E_{\overline{\mathrm{M}}/2}^{uudd}= E_{\overline{\mathrm{M}}/2}^{uudd} - E_{\overline{\mathrm{M}}/2}^{SS}  \label{eq4b} \\
\Delta E_{3\overline{\mathrm{K}}/4}^{uudd}= E_{3{\overline{\mathrm{K}}}/4}^{uudd} - E_{3{\overline{\mathrm{K}}}/4}^{SS} \label{eq4c} \\
\nonumber
\end{gather} 
see Fig.~\ref{fig:f2} and Supplementary Table 2 for energy differences.

Upon taking the higher-order exchange constants into account, we need to modify the first three Heisenberg pair-wise exchange interaction parameters obtained from fitting of the homogeneous spin spirals neglecting higher-order interactions as follows (for a derivation see Ref.~\cite{hoi2020}):

\begin{subequations} \label{eq:modJs}
	\begin{align}	
	J^{\prime}_{1}&= J_{1}-Y_{1} \\
	J^{\prime}_{2}&= J_{2}-Y_{1} \\
	J^{\prime}_{3}&= J_{3}-B_{1}/2 
	\end{align}
\end{subequations}

Note that the four-site four spin interaction does not adjust any exchange interaction parameters, since its contribution to the homogeneous spin spirals is a constant term of $-$12$K_{1}$.

Due to the large 2D unit cell, the energy of the multi-$Q$ states were evaluated from asymmetric films consisting of 8 layers in total. However, we have checked for zero electric field that the energy difference between the 8 and 11 layer film calculations is less than 1 meV.\\

\noindent{\textbf{Atomistic spin dynamics simulations.}}
To relax and calculate the energy of the spin spirals, skyrmion lattice, FM phases as well as the isolated skyrmions, we  
performed spin dynamics simulations based on
the Landau-Lifshitz equation: 

\begin{gather} \label{eq:llg}
\hslash \frac{d\textbf{m}_{i}}{dt}=\frac{\partial H}{\partial{\textbf{m}}_{i}}\times\textbf{m}_{i}-\alpha\left(\frac{\partial H}{\partial{\textbf{m}}_{i}}\times\textbf{m}_{i}\right)\times\textbf{m}_{i}
\end{gather}

where $\hslash$ is the reduced Planck constant, $\alpha$ is the damping parameter and the Hamiltonian $H$ is defined in Eq. (1). We used a time step of 0.1~fs, $\alpha$ is varied from 0.05 to 0.1 and the simulations were carried out over 4 to 6 million steps for relaxation. We solve Eq.~(\ref{eq:llg}) by semi-implicit method as proposed by Mentink $\textit {et al}$. \cite{mentink2000}.\\

\noindent{\textbf{Geodesic nudged elastic band method.}}
We first create isolated skyrmions in the field-polarized background, i.e., at a magnetic field above $B_{\mathrm c}$ from the theoretical profile \cite{Bogdanov1994} and then relax the spin structure using spin dynamics with the full set of DFT parameters (Table I). We computed the collapse and creation barriers of isolated skyrmions using the GNEB method \cite{gneb}. The method finds the minimum energy path (MEP) connecting the initial state (IS), i.e., skyrmions, and final state (FS), i.e., FM state, on a multidimensional energy surface. Within GNEB, an initial path connecting the IS and FS is created by a chain of images of the system. The objective of the method is to bring the initial path to MEP via relaxing the intermediate images. The relaxation is achieved by a force projection scheme. For this, the effective field is calculated at each images and its local tangent to the path is replaced by a spring force which maintains a uniform distribution of images. The maximum energy of MEP corresponds to the saddle point (SP) which defines the energy barrier separating two stable states. We compute the energy of the SP accurately using a climbing image (CI) method on top of GNEB.\\

\noindent{\large{\textbf{Data availability}}}

The authors declare that the data supporting the findings of this
study are available within the article and its Supplementary Information files.\\

\noindent{\large{\textbf{Code availability}}}

The atomistic spin dynamics code used in this work is available from the authors upon a reasonable request. 

\bibliographystyle{naturemag}
\bibliography{reference}

\begin{thebibliography}{10}
\bibitem{footnote1}
\bibinfo{title}{Note that the spin spiral periods obtained at zero magnetic
  field in the atomistic spin dynamics simulations are consistent with those
  from the {DFT} calculations.}
\bibitem{footnote2}
\bibinfo{title}{Note that the four-spin term exhibits the same field dependence
  for creation and collapse barriers since it has a similar contribution to the
  fm state as to the skyrmion state.}
\end{thebibliography}

\vspace{0.4cm}

\noindent{\large{\textbf{Acknowledgements}}}\par

We gratefully acknowledge computing time at the supercomputer of the North-German Supercomputing Alliance (HLRN) and financial support from the Deutsche Forschungsgemeinschaft (DFG, German Research Foundation) via project no.~414321830 (HE3292/11-1). S.P. graciously acknowledge financial support from the European Research Council (ERC) under the European Union's Horizon 2020 research and innovation program (Grant No. 856538, project ``3D MAGiC". S.P. and S.H. thank G. Bihlmayer and M. Hoffmann for useful discussion. S.H. thanks M. Goerzen for insightful discussions.\\

\noindent{\large{\textbf{Author contributions}}}\par
S.H. devised the project. S.P. performed the calculations. S.P. and S.H. analyzed the results and wrote the paper.\\ 

\noindent{\large{\textbf{Competing interests}}}\par

The authors declare no competing financial interests.\\

\noindent{\large{\textbf{Additional information}}}\par 

Supplementary information is available for this paper at
 
\end{document}